\documentclass[journal]{IEEEtran}
\usepackage{amsmath,amsfonts}
\usepackage{algorithmic}
\usepackage{array}
\usepackage[caption=false,font=normalsize,labelfont=sf,textfont=sf]{subfig}
\usepackage{textcomp}
\usepackage{stfloats}
\usepackage{url}
\usepackage{verbatim}
\usepackage{graphicx}
\usepackage{booktabs}
\def\BibTeX{{\rm B\kern-.05em{\sc i\kern-.025em b}\kern-.08em
		T\kern-.1667em\lower.7ex\hbox{E}\kern-.125emX}}
\usepackage{balance}
\newcounter{MYtempeqncnt}
\usepackage{caption}
\begin{document}
	\title{Energy Consumption Analysis for Continuous Phase Modulation in Smart-Grid Internet of Things of beyond 5G}
	\author{Hongjian Gao, Yang Lu, Shaoshi Yang, Jingsheng Tan, Longlong Nie and Xinyi Qu 
		\thanks{\textit{Corresponding author: S. Yang (E-mail: shaoshi.yang@bupt.edu.cn)} 
		
			H. Gao and Y. Lu are with the State Grid Smart Grid Research Institute Co., Ltd., Beijing 102209, China.
			
			S. Yang, J. Tan, L. Nie, and X. Qu are with the School of Information and Communication Engineering, Beijing University of Posts and Telecommunications, and also with the Key Laboratory of Universal Wireless Communications, Ministry of Education, Beijing 100876, China.
			
			Published on \textit{Sensors}, vol. 24, no. 2, article number 533, Jan. 2024,  https://doi.org/10.3390/s24020533
	}}
	
	\markboth{
	}%
	{How to Use the IEEEtran \LaTeX \ Templates}
	
	\maketitle

\begin{abstract}
Wireless sensor network (WSN) underpinning the smart-grid Internet of Things (SG-IoT) has been a popular research topic in recent years due to its great potential for enabling a wide range of important applications. However, the energy consumption (EC) characteristic of sensor nodes is a key factor that affects the operational performance (e.g., lifetime of sensors) and the total cost of ownership of WSNs. In this paper, to find the modulation techniques suitable for WSNs, we investigate the EC characteristic of continuous phase modulation (CPM), which is an attractive modulation scheme candidate for WSNs because of its constant envelope property. We first develop an EC model for the sensor nodes of WSNs by considering the circuits and a typical communication protocol that relies on automatic repeat request (ARQ)-based retransmissions to ensure successful data delivery. Then, we use this model to analyze the EC characteristic of CPM under various configurations of modulation parameters. Furthermore, we compare the EC characteristic of CPM with that of other representative modulation schemes, such as offset quadrature phase-shift keying (OQPSK) and quadrature amplitude modulation (QAM), which are commonly used in communication protocols of WSNs. Our analysis and simulation results provide insights into the EC characteristics of multiple modulation schemes in the context of WSNs; thus, they are beneficial for designing energy-efficient SG-IoT in the beyond-5G (B5G) and the 6G era.
\end{abstract}


\section{Introduction}
\IEEEPARstart{S}{mart} grid is the energy infrastructure for smart cities, telecommunications, networks, and the computing industry. It upgrades traditional power grid systems with state-of-the-art information and communication technologies, such as wireless sensor network (WSN) techniques and the Internet of Things particularly designed for the smart grid industry (SG-IoT). In fact, SG-IoT heavily relies on WSN, which is characterized by a variety of distinct performance metrics, such as transmission rate, signal coverage, energy consumption (EC), and~network lifetime~\cite{7570253}. 
(defined as the number of joules consumed per successfully transmitted bit) of wireless sensors, because the energy supply requirements of sensors are stringent in WSNs (i.e., very limited energy supply) and all the other performance metrics can be affected by the EC characteristic of sensors.  To elaborate a little further, in~many application scenarios of the smart grid industry, it is often inconvenient or unsafe for humans to work in the deployment site and~the lifetime of sensors is often expected to be over several years. Therefore, low EC is of vital importance for these scenarios.

By contrast, in wireless communication systems that operate with the support of power grid infrastructure, it is more appropriate to invoke energy efficiency (EE), which is typically defined as the number of bits successfully transmitted per joule. This concept is at the heart of green communications, a~vision globally recognized for reducing the Carbon footprint produced by the networking sector, especially in the era of 5G, 5G-Advanced, and 6G \cite{6G_slice}. Obviously, EE is the reciprocal of EC. Extensive studies have been devoted to optimizing the EE of wireless networks in the past decade. For~instance, in~\cite{Kent_EE_TCOM}, based on the fractional programming framework, the~joint power and subcarrier allocation problem was solved for maximizing the EE of a multi-user, multi-relay, single-cell orthogonal frequency-division multiple access (OFDMA) cellular network composed of single-antenna nodes. 
Afterwards, system models that are more complicated were considered: the joint transmit and receive beamforming-based multi-user, multi-relay, multi-input multi-output (MIMO)-OFDMA cellular networks~\cite{Kent_GC_EE,Kent_TWC}; the~multi-cell single-antenna OFDMA networks~\cite{EE_Wenpeng}; the~partial/full interference alignment-based multi-user, multi-relay, multi-cell MIMO-OFDMA networks~\cite{Kent_TSP}; the~massive MIMO-aided, multi-pair, one-way decode-and-forward relay system~\cite{EE_MP_DF_mMIMO}; and the~fully connected $K$-user interference channel with each user having either a single antenna or multiple antennas~\cite{EE_IA}. Additionally, the~EE of wireless networks that are delay-sensitive was also studied by maintaining statistical quality-of-service {QoS} guarantees in OFDMA networks~\cite{EE_effective_capacity} and by considering the uplink ultra-reliable low-latency communication (URLLC) traffic in the MIMO-aided grant-free access~\cite{EE_MIMO_URLLC_grant_free} of 5G and its beyond. In~\cite{secrecy_EE}, secrecy-energy efficient hybrid beamforming schemes were designed for a satellite-terrestrial integrated network in~order to maximize the achievable secrecy-EE while satisfying the signal-to-interference-plus-noise ratio (SINR) constraints of both the earth stations and the cellular users; further,~in~\cite{SLNR_secure_EE_beam_satellite}, the secrecy-energy efficient beamforming in multibeam satellite systems was investigated with the metric of signal-to-leakage-plus-noise ratio (SLNR).  

Since wireless sensors are typically powered by batteries, it is more appropriate to use EC in the context of wireless sensors. EC is closely related to the selected modulation scheme. Firstly, this selection may influence the type of electronic components utilized, such as a power amplifier (PA) or analog-to-digital converter (ADC), because~different modulation schemes may require different circuit designs and implementations. Secondly, the~specific choice of modulation schemes also affects the number of bits transmitted in a single symbol duration that consumes a certain amount of energy. Thirdly, different modulation schemes may incur different packet error rates (PERs), which influence the number of retransmissions that also consume energy and are necessary for successful packet delivery between any pair of wireless sensors. Therefore, it is important to investigate the impact of different modulation schemes on the EC and identify the most appropriate scheme for WSNs of SG-IoT.

Prior research mainly focused on studying the EC of modulation schemes that are sensitive to the nonlinearity of PAs. More specifically, in~\cite{cui2005energy}, the~EC minimization problems corresponding to $M$-ary quadrature amplitude modulation (MQAM) and multiple frequency-shift keying (MFSK) were studied. In \cite{6240051, Wang2008MinimizationOT}, the~authors studied the relationship between the total EC per successfully transmitted information bit and the transmission distance while assuming different modulation methods, such as binary phase-shift keying (BPSK), quadrature phase-shift keying (QPSK), and~16QAM. 
They also studied the average signal-to-noise ratio (SNR) values required to achieve the optimal EC. In~\cite{Abo}, the transmission power of MQAM was optimized by using a particular model to achieve the minimum EC. In~\cite{6327311}, the EC per successfully transmitted bit for modulation techniques including binary frequency-shift keying (BFSK), BPSK, QPSK, 16QAM, and 64QAM was studied under various channel conditions. 
However, these modulation techniques require the use of linear PAs, which results in lower energy utilization efficiency. In~contrast, constant envelope modulation techniques are insensitive to the nonlinearity of PAs; thus, they constitute a promising solution to improving the energy utilization efficiency. However, there is a scarcity of research focusing on the impact of constant envelope modulation techniques on the achievable EC in the context of WSNs underpinning SG-IoT.

Against the above backdrop, in~this paper, we endeavor to investigate the impact of a constant envelope modulation technique, i.e.,~continuous phase modulation (CPM), on~the EC of sensor nodes in WSNs suitable for SG-IoT. Our novel contributions are summarized as~follows. 

\begin{itemize}
	\item{We establish a realistic power consumption model through the analysis of circuit power consumption,~transmission power consumption, and~reception power consumption on a point-to-point communication link; in~particular, we consider three operation modes of the sensors, including sleeping mode,~transient mode, and~active mode.  }
	\item{Based on the above power consumption model and a typical automatic repeat request (ARQ)-based wireless transmission protocol, the~EC incurred by successfully sending a single information bit is numerically evaluated under different configurations of CPM parameter values. In~particular, we consider different waveform pulses of the CPM, including the rectangular pulse,~rising cosine pulse, and~GMSK pulse, for comprehensive coverage. We also investigate the impact of the distance between the transmitter and the receiver, the~impact of the received SNR,~the impact of the modulation order, and the average number of transmissions required for sending a single packet, under various modulation schemes considered.}
	\item{We compare the EC per successfully transmitted bit of the CPM with that of conventional non-constant envelope modulation methods, such as offset quadrature phase-shift keying (OQPSK) used in the Zigbee standard and QAM modulation supported by the current 5G standard. Our simulation results and analysis demonstrate that CPM enjoys a significantly lower EC than OQPSK and 16QAM in the scenario considered, which is valuable for the standard evolution of beyond 5G tailored for the important use case of low-power SG-IoT.}
\end{itemize}

\section{The EC~Model}\label{Sec_2}
To analytically determine the amount of energy consumed when a single bit is transmitted without error, an~EC model needs to be established. We make the assumption that each packet transmitted in the forward direction induces an error-free feedback packet in the reverse direction, which acknowledges the successful reception of the data packet or requests for retransmission.

\subsection{Packet~Structure}
\begin{figure}[tbp]
	\centering
	\includegraphics[width=3.5in]{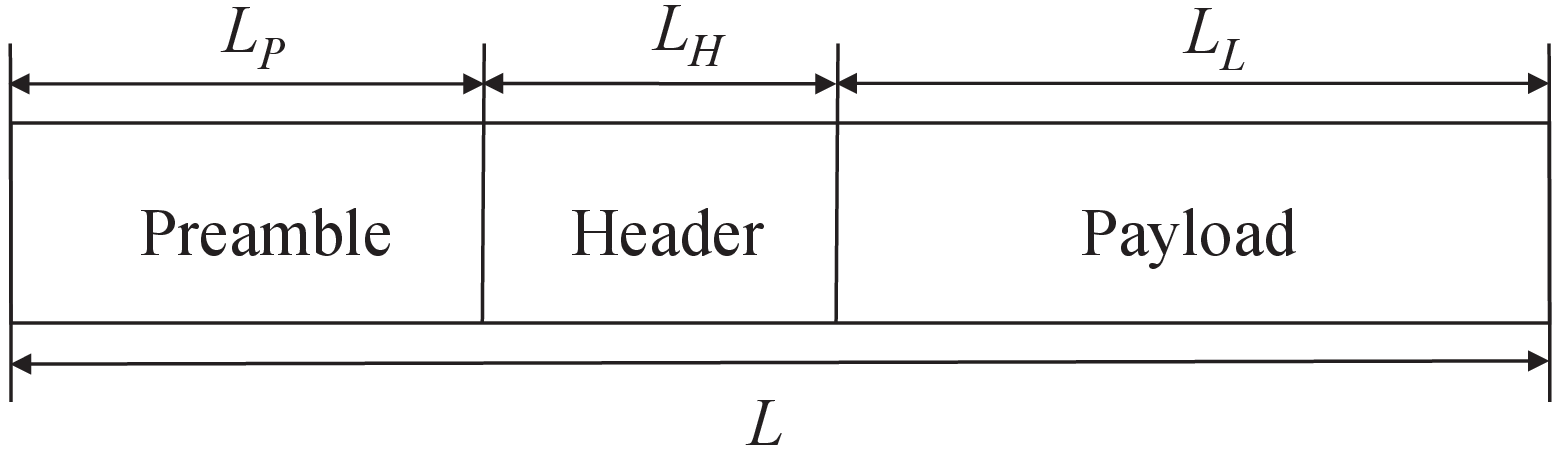}
	\caption{Physical layer packet~structure.}
	\label{fig1}
\end{figure}

In wireless communication systems, the~general format of the physical layer packet structure is shown in Figure~\ref{fig1} and consists of three parts: a pilot code for clock synchronization, a~packet header specifying the configuration of transmission parameters, and~a data payload carrying the transmitted~data. 

We assume that the entire packet uses the same modulation and~that the symbol error probability (SEP) is determined by the received SNR $\gamma$ and the modulation scheme adopted. We also assume that symbol errors are independently and identically distributed (i.i.d.) and no channel coding is used, i.e.,~no redundant bits are added (redundant bits increase EC). If~a symbol is erroneously detected at the receiver, the~entire packet will be retransmitted until all the symbols in the packet are correctly detected at the receiver. The~packet error probability (PEP) can be expressed as a function of the SEP, the~packet length $L$, and~the number of bits per symbol $m=\log_2 M $, i.e.,
\begin{equation}
	\mathsf{PEP}=1-(1-\mathsf{SEP})^{L/m}, 
\end{equation}
where $M$ denotes the size of the modulation constellation.

Hence, by~utilizing the packet length $L$ and the probability that a packet is successfully transmitted, i.e.,~$1-\mathsf{PEP}$, the~average amount of data successfully delivered per transmission duration of a packet is formulated as
\begin{equation}
	N_c=L(1-\mathsf{SEP})^{L/m}.
\end{equation}
Then, upon assuming that the feedback signal indicating whether a retransmission is needed or not is reliably transmitted on the reverse link, the average number of transmissions required for successfully delivering a packet is given by 
\begin{equation}
	N_\textrm{re} = \frac{L}{N_c} = \frac{1}{1-\mathsf{PEP}} = \frac{1}{(1-\mathsf{SEP})^{L/m}}.
\end{equation} 

\subsection{Basics of~CPM}\label{subsec:basics_CPM}
CPM is an attractive modulation scheme for WSNs underpinning SG-IoT because~its carrier phase is modulated in a continuous manner and~it is typically implemented as a constant-envelope waveform, i.e.,~the transmitted carrier power is constant.  The~phase continuity requires a relatively small percentage of the power to occur outside of the intended band (e.g., low fractional out-of-band power), leading to high spectral efficiency. Meanwhile, the~constant envelope yields excellent power/energy efficiency. However, the~primary drawback of CPM is the high implementation complexity required for an optimal~receiver.

For systems that employ CPM, the~transmitted signal at time instant $t$ can be expressed as\cite{1095001}
\begin{equation}
	s(t,\boldsymbol{I})=\sqrt{\frac{2E}{T}}\cos({2\pi f_c t+\phi(t,\boldsymbol{I})+\phi_0}),
\end{equation}
where $E$ is the symbol energy, $T$ is the symbol interval, $f_c$ is the carrier frequency, and~$\phi_0$ is an arbitrary constant initial phase shift that can be set to zero without loss of generality when coherent transmission is considered.  
In addition, $\phi(t,\boldsymbol{I})$ 
is the time-varying information-carrying phase formulated as
\begin{equation}
	\phi(t,\boldsymbol{I})=2\pi \sum_{k=-\infty}^{K}h_k I_k q(t-kT), KT \leq t \leq (K+1)T,
\end{equation}
where $\boldsymbol{I} = \{ I_k|k \in (-\infty,\cdots,-1,0,+1,\cdots, K) \}$ is an infinitely long sequence of uncorrelated $M$-ary data symbols, each having one of the values from the alphabet\linebreak  $\mathcal{A}=\{\pm 1, \pm 3, \cdots, \pm (M-1)\}$ with equal probability $1/M$; $\{h_k\}$ is a sequence of modulation indices defined as $h_k = 2f_{d,k} T$, with~$f_{d,k}$ being the peak frequency deviation. When $h_k = h$ for all $k$, the~modulation index remains fixed for all symbols. When the modulation index changes from one symbol to another, the~signal is called multi-$h$ CPM, with~$h_k$ varying in a cyclic manner. $q(t)$ is some normalized waveform shape that represents the baseband phase response (i.e., phase pulse) and is obtained from the frequency pulse $g(t)$ by
\begin{equation}
	q(t) = \int_{-\infty}^t g(\tau)d\tau. 
\end{equation} 

If the duration of $g(t)$ is equal to the symbol interval $T$, namely, $g(t) =0$ for $t>T$, the~modulated signal is called full-response CPM. If~the duration of $g(t)$ is larger than the symbol interval $T$, namely, $g(t) \neq 0$ for $t >T$, the~modulated signal is called partial-response~CPM. 

Suppose the length of the frequency pulse $g(t)$ in terms of the number of symbol intervals is $N$. Thus, $N=1$ yields full-response CPM. If~$g(t)$ is selected as a rectangular pulse, namely,
	\begin{equation}
		g(t)=
		\begin{cases}
			\frac{1}{2NT},& \text{$ 0\leq t \leq NT $},\\
			0,& \text{otherwise},
		\end{cases}
	\end{equation}
	then for 
	a full-response CPM, we have
	\begin{equation}
		q(t)=
		\begin{cases}
			0, & \text{$t \leq 0 $},\\
			\frac{t}{2T},& \text{$ 0\leq t \leq T $},\\
			\frac{1}{2},& \text{$t \geq T$}.
		\end{cases}
	\end{equation}

It is evident that the performance of CPM is influenced by certain parameters, including but not limited to $M$, $h_k$, $N$, and~the frequency pulse $g(t)$. 
Note that by choosing different pulse shapes $g(t)$ and varying $M$, $h_k$, and $N$, an~infinite variety of CPM signals may be generated, each with its unique characteristics and~performance.

For a CPM signal, the~error rate performance can be derived based on the maximum-likelihood sequence detection (MLSD) receiver, which is conventionally computed using the Viterbi Algorithm (VA). Specifically, for a given CPM scheme, we have
\begin{equation} \label{SEP}
	\mathsf{SEP}=K_{\textrm{min}}Q(\sqrt{d_{\textrm{min}}^2 \gamma}).
\end{equation}

According to Anderson's seminal work on digital phase modulation \cite{anderson2013digital}, $K_{\textrm{min}}$ denotes the total number of feasible paths that satisfy the constraint of the minimum Euclidean distance $d_{\textrm{min}}$ within the observation interval on the CPM phase grid. The value of $K_{\textrm{min}}$ increases with  the modulation order $M$. Both $K_{\textrm{min}}$ and $d_{\textrm{min}}$ depend on critical parameters including $M$, $h_k$, $N$, and~the pulse shaping function $g(t)$.

\subsection{Circuit Power~Consumption}

In wireless communication systems, a~significant portion of energy is dedicated to signal transmission and reception circuits, which are mainly composed of the baseband (BB) digital signal processing unit and the radio frequency (RF) signal processing unit, as~shown in Figure~\ref{entire_power_model}. To~elaborate a little further, the~BB signal processing unit mainly includes source coding/decoding, pulse shaping, channel coding/decoding, digital modulation/demodulation, channel estimation, synchronization, and~so on.
For a wireless sensor, since the data rate requirement is usually low, the~BB symbol rate is also low. Meanwhile, typically, no computation-intensive signal processing techniques, such as multi-user detection and iterative decoding, are used in an energy-constrained wireless sensor; hence, the BB power consumption is significantly smaller than the RF circuit power consumption.

A typical model of an RF signal processing unit, also known as an RF chain, is shown in Figure~\ref{RF_power_model} \cite{cui2005energy, Wang2008MinimizationOT, 4068131, Wang2008, Shunqing_2020}. Specifically, on~the transmitter side, the~BB signal is first converted to an analog signal by the digital-to-analog converter (DAC). Then, the analog signal is filtered by the low-pass filter and upconverted by the mixer, whose output is then filtered again, amplified by the power amplifier (PA), passed through the duplexer, and~finally transmitted to the wireless channel. On~the receiver side, the~RF signal is sequentially filtered, amplified by the low-noise amplifier (LNA), cleaned by the anti-aliasing filter, downconverted by the mixer, filtered again before passing through the intermediate frequency amplifier (IFA) that has an adjustable gain, and~finally converted back to a digital signal by the analog-to-digital converter (ADC). Note that the mixers operate with the aid of the local oscillator (LO) and,~among all the RF components, the~PA and LNA usually have much higher power consumption than the others.

\begin{figure}[tbp]
	\centering
	\includegraphics[width=3.5in]{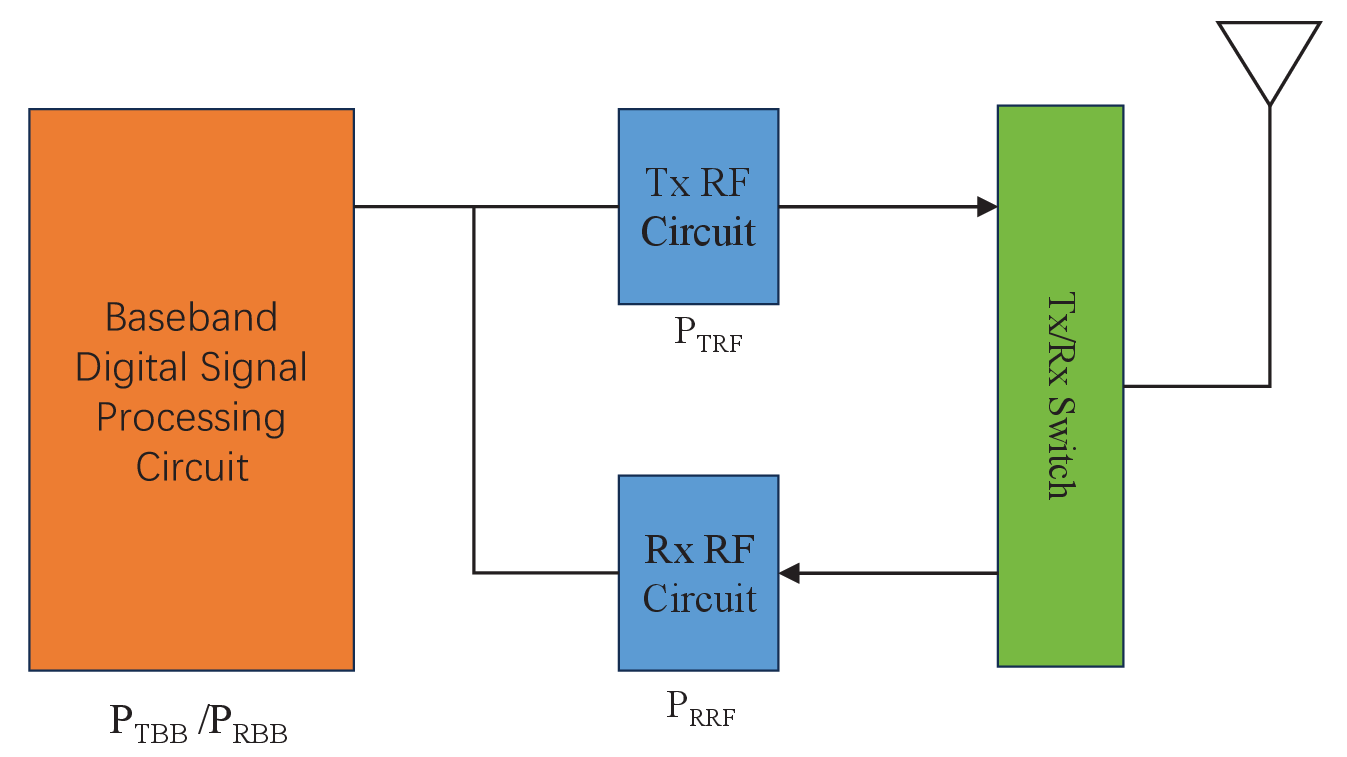}
	\caption{The communication modules and the corresponding power consumption model of a point-to-point wireless communication system.}
	\label{fig1entire_power_model}
\end{figure}

\begin{figure}[tbp]
	\centering
	\includegraphics[width=3.5in]{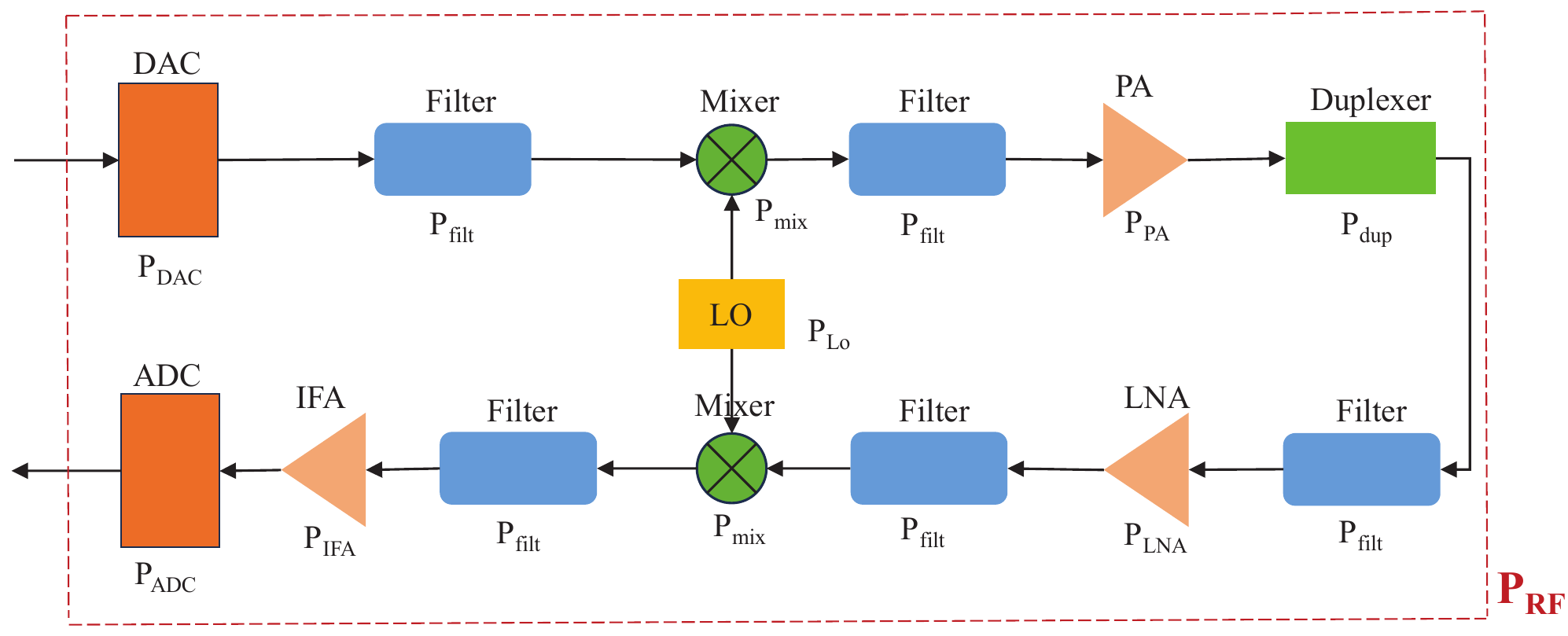}
	\caption{A typical power consumption model of the RF signal processing unit of a point-to-point wireless communication system.}
	\label{RF_power_model}
\end{figure}

Accordingly, the~total power consumption during transmission can be expressed as
\begin{equation}
		P_{\textrm{Tx}}(d)=P_{\textrm{TBB}}+\underbrace{ P_{\underline{\textrm{TRF}}}+ P_{\textrm{PA}}(d)}_{P_{\textrm{TRF}}}=P_\textrm{T0}+\frac{\xi}{\eta}P_{\textrm{T}}(d),
\end{equation}
where $P_{\textrm{TBB}}$ and $P_{\underline{\textrm{TRF}}}$ represent the power consumption of the transmitter's BB processing unit and RF processing unit excluding the PA, respectively, and~both of them can be regarded as constant values that are collectively denoted by $P_\textrm{T0}$; $P_{\textrm{PA}}(d)$ and $P_\textrm{T}(d)$ are defined as the power consumption of the PA and the transmit power, respectively, both of which are functions of the transmission distance $d$ upon assuming adaptive power control and are related via $P_{\textrm{PA}}(d) = \frac{\xi}{\eta}P_{\textrm{T}}(d)$; and $\eta$ and $\xi$ represent the drain efficiency and the peak-to-average power ratio (PAPR) of the PA, respectively~\cite{6327311}. 

Similarly, the~total power consumption of the receiver is expressed as
\begin{equation} 
		P_{\textrm{Rx}}=P_{\textrm{RBB}}+ \underbrace{ P_{\underline{\textrm{RRF}}}+P_{\textrm{LNA}}}_{P_{\textrm{RRF}}}=P_{\textrm{R0}},
\end{equation}
where $P_{\textrm{RBB}}$ and $P_{\underline{\textrm{RRF}}}$ represent the power consumption of the receiver's BB processing unit and RF processing unit excluding LNA, respectively, and~both of them can be regarded as constant values; $P_{\textrm{LNA}}$ represents the power consumption of the LNA, which is also constant upon assuming that the LNA is appropriately designed and biased, so that necessary sensitivity is provided for reliably receiving, demodulating, and decoding a minimum power signal. Hence, $P_{\textrm{RBB}}$, $P_{\underline{\textrm{RRF}}}$, and~$P_{\textrm{LNA}}$ are collectively denoted by the constant $P_{\textrm{R0}}$.
	
\subsection{Transmission Power~Consumption}

Due to the path-loss, scattering, reflection, and other phenomena in the wireless channel, a~certain amount of energy is inevitably lost during the transmission of electromagnetic waves that carry data symbols, thus resulting in transmission energy dissipation. The~path-loss depends on several factors, such as the distance between the transmitter and receiver, the~frequency of the signal, the~type of antennas used, and~the environment through which the signal~propagates.

When the signal with a transmit power $P_\textrm{T}(d)$ is propagated through the wireless channel, the~received power $P_\textrm{R}(d)$ at the receiver can be formulated as
\begin{equation} \label{P_R}
	P_\textrm{R}(d)= G_\textrm{T}G_\textrm{R}P_\textrm{T}(d)\Big(\frac{\lambda}{4\pi d}\Big)^\alpha = \frac{P_\textrm{T}(d)}{A_0d^\alpha},
\end{equation}
according to Friis' transmission equation. This expression characterizes the dependency between the received power with respect to several parameters. Specifically,  the constant $A_0$ depends on the transmit antenna gain $G_\textrm{T}$, the~receive antenna gain $G_\textrm{R}$, and~the carrier wavelength $\lambda$. Additionally, the~path-loss exponent is $\alpha$. It is noted that the received power is inversely proportional to the distance $d$ raised to the power of $\alpha$, and~the received SNR $\gamma$ is proportional to $1/A_0$ divided by $d^\alpha$. These relationships collectively describe how the aforementioned parameters influence the received power and the received SNR~\cite{6327311}:
\begin{equation}  \label{Receive_SNR}
	\gamma=\frac{P_\textrm{R}(d)}{N_0WN_\textrm{f}M_\textrm{l}}, 
\end{equation}
where $W$ represents the transmission bandwidth, $N_0$ denotes the power spectral density of the baseband-equivalent additive white Gaussian noise (AWGN), and~they are primary determinants of $\gamma$. Furthermore, the~noise figure of the RF front-end of the receiver, denoted as $N_\textrm{f}$, and~any additional noise or interference, represented by the link margin term $M_\textrm{l}$, can also impact the received SNR $\gamma$. 

Accordingly, by~substituting Equation~(\ref{P_R}) into Equation~(\ref{Receive_SNR}), the~relationship between the transmit power $P_\textrm{T}(d)$, the~communication distance $d$, the~received SNR $\gamma$, and~other parameters can be quantitatively expressed as
\begin{equation}    
	P_\textrm{T}(d)=A_0d^\alpha N_0WN_\textrm{f}M_\textrm{l}\gamma =Ad^\alpha \gamma,
\end{equation}
where we have $A=A_0 N_0W N_\textrm{f}M_\textrm{l}$.

\subsection{EC per Successfully Transmitted~Bit}
As mentioned in Section~\ref{Sec_2}, we assume that each packet transmitted in the forward direction is matched by an error-free feedback packet in the reverse direction in~order to guarantee reliable transmission. Both directions of transmissions consume energy.  The~above transmission process, usually incorporating retransmissions, continues until the entire packet of the forward direction is correctly decoded at the receiver. Additionally, we assume that the sensor node transceiver circuitry works in a multi-mode manner: (1)~when there are data to transmit or receive, all circuits of the sensor work in the active mode; (2)~when neither transmission nor reception are needed, the~circuits of the sensor enter sleep mode by default, which uses the minimum possible power (small enough to be negligible) to ensure that the circuits can be activated when necessary; (3) when the sensor is in the period of switching from sleep mode to active mode, it is in a transient mode that also consumes non-negligible energy. Note that the transient duration from active mode to sleep mode is sufficiently short to be neglected but~the start-up process from sleep mode to active mode can be slow.

Therefore, for~a single round-trip transmission (forward direction transmission and reverse direction feedback) on a point-to-point communication link, the~total EC of the system can be divided into two  parts: the EC of the forward direction transmission and the EC of the reverse direction feedback. For~the forward direction transmission, the~EC is composed of the start-up energy consumption $2E_\textrm{ST}$ in transient mode (both the transmitting node and the receiving node may be in sleep mode initially), the~transmitter energy consumption $E_\textrm{Tx}(d)$ in active mode, and~the receiver energy consumption $E_\textrm{Rx}$ in active mode. Therefore, the~EC of the forward direction transmission is expressed as
	\begin{equation}
		E_{\textrm{FW}}=2E_{\textrm{ST}}+ E_\textrm{Tx}(d) + E_\textrm{Rx}   =2 E_{\textrm{ST}} + P_{\textrm{Tx}}(d)T_\textrm{DTA}+P_{\textrm{Rx}}T_\textrm{DRA},
	\end{equation}
	where $T_\textrm{DTA}$ is the transmission duration for sending a data packet in the forward direction and~$T_\textrm{DRA}$ is the corresponding duration of signal processing at the receiver of the forward~direction. 

Additionally, the~EC of the reverse direction transmission is expressed as
	\begin{equation}
		E_{\textrm{RV}} = \underline{P}_{\textrm{Tx}}(d) T_\textrm{FTA}+ P_{\textrm{Rx}}T_{\textrm{FRA}},
	\end{equation}
	where $\underline{P}_{\textrm{Tx}}(d)$ is the total power consumption for transmission of a feedback packet in the reverse direction by the receiving node of the forward direction and~$T_\textrm{FTA}$ is the corresponding time duration. Note that $\underline{P}_{\textrm{Tx}}(d)$ may be different from $P_{\textrm{Tx}}(d)$ of the forward direction because~a feedback packet may have a different transmission rate and reliability requirements compared with a data packet.  $T_\textrm{FRA}$ is the duration of processing the feedback packet at its receiver (i.e., the~transmitting node of the forward direction) in active mode.

Based on the above analysis, we obtain the EC per successfully transmitted bit as \eqref{eq:EC_per_bit}, 
\begin{figure*}[!t]
	\centering
	\hrulefill
	\vspace*{8pt}
	\begin{equation}\label{eq:EC_per_bit}
		\begin{aligned}
			E_b&=\frac{N_\textrm{re}(E_{\textrm{FW}}+E_{\textrm{RV}})}{L}\\
			& = \frac{E_{\textrm{FW}}+E_{\textrm{RV}}}{N_c}\\
			&=\frac{2E_{\textrm{ST}}+P_{\textrm{Tx}}(d)T_\textrm{DTA}+P_{\textrm{Rx}}T_\textrm{DRA}+\underline{P}_{\textrm{Tx}}(d) T_\textrm{FTA}+ P_{\textrm{Rx}}T_{\textrm{FRA}}}{L(1-\mathsf{SEP})^{L/m}}\\
			&=\frac{2E_{\textrm{ST}}+(P_\textrm{T0}+\frac{\xi}{\eta}P_{\textrm{T}}(d))T_\textrm{DTA} + P_{\textrm{Rx}}T_\textrm{DRA} + (P_\textrm{T0}+\frac{\xi}{\eta}\underline{P}_{\textrm{T}}(d))T_\textrm{FTA} + P_{\textrm{Rx}}T_{\textrm{FRA}}}{L(1-K_{\textrm{min}}Q(\sqrt{d_{\textrm{min}}^2 \gamma}))^{L/m}}.
		\end{aligned}
	\end{equation}
\end{figure*}
where $\underline{P}_{\textrm{T}}(d)$ is the transmit power for the feedback packet transmission and~$N_\textrm{re}$ is the average number of retransmissions. 

\section{CPM Parameter~Selection}
As mentioned in Section~\ref{subsec:basics_CPM}, the~main parameters that affect the  performance of CPM are $M$, $h_k$, $N$, and the frequency pulse shaping function $g(t)$, whilst the minimum squared Euclidean distance $d_\textrm{min}^2$ also depends on these parameters. Therefore, it is necessary to determine how these parameters influence the EC per successfully transmitted bit.

In principle, it is possible to implement an infinite number of different CPM signals using various combinations of design parameters. However, for practical implementation, we must consider the trade-off between the achievable performance and the cost incurred. It is well known that partial-response CPM signals usually have better spectral efficiency than full-response CPM signals. However, the computational complexity of the optimal MLSD receiver exponentially increases with $N$, which is the~length of $g(t)$ in terms of the number of symbol intervals. In~this study, we focus on the partial-response CPM with a moderate value of $ N = 3$. The~modulation order $M$ also significantly influences the computational complexity and the requirements for demodulation devices \cite{Aulin_partial_response}, which can lead to high EC. Therefore, small values, such as $M = 2$, $4$, $8$, $16$, etc., are generally chosen. We assume the modulation index $h_k = h$, which is also an important parameter and has a complex functional relationship with the minimum squared Euclidean distance \mbox{$d_\textrm{min}^2$ \cite{Aulin_partial_response,  SSB_CPM}.} Smaller values of $h$ result in narrower bandwidth, more concentrated signal energy, and~narrower transition bands. However, they also make the phase variations less obvious and can increase the complexity of demodulation decisions. For~single index modulation, $h = 0.5$ or $0.75$ is commonly used. Finally, the pulse shaping function $g(t)$ usually utilizes rectangular pulse (REC), rising cosine pulse (RC), and~Gauss minimum-phase shift-keying pulse (GMSK) \cite{6956787}.

Given the imperative to minimize EC for sensors, the~present study focuses on CPM signals that can be implemented with simple devices and require low computational complexity. Accordingly, CPM signals with three different $g(t)$ functions are selected for investigation, assuming $M=2, 4, 8, 16$, $h=0.75$, and $N=3$. 
The~values of $d_{\textrm{min}}^2$ under these parameter configurations are calculated with the methods given in \cite{Aulin_partial_response, SSB_CPM} and listed in Table~\ref{tab1}.

\begin{table}
	\begin{center}
		\caption{The value of $d_{\textrm{min}}^2$ when we set $M=2, 4, 8, 16$, $h= 0.75$, and $N = 3$. For GMSK, the time-bandwidth product $BT$ is set to 0.3, where $B$ is the $-3$ dB bandwidth of the Gaussian pulse.}
		\label{tab1}
		\begin{tabular}{ c | c | c |c|c}
			\hline
			Waveform & $M=2$ & $M=4$ & $M=8$ & $M=16$ \\
			\hline
			REC & 2.31648 & 1.41550 & 2.12325 & 2.831  \\
			RC & 2.96059 & 5.30037 & 6.12447 & 8.16596 \\ 
			GMSK & 2.89955 & 4.69275 & 5.95011 & 7.93348 \\ 
			\hline 
		\end{tabular}
	\end{center}
\end{table}

\section{Evaluation the EC of~CPM}
\unskip

\subsection{Identification of Major Performance Influencing Factors}
From Equation~(\ref{eq:EC_per_bit}), it can be observed that the EC per successfully transmitted bit, $E_b$, is predominantly determined by four parameters: the forward-link transmit power $P_{\textrm{T}}(d)$, the reverse-link transmit power $\underline{P}_{\textrm{T}}(d)$, the specific CPM scheme adopted, and the received SNR $\gamma$. From another perspective, we can also say that $E_b$ is mainly determined by  $P_{\textrm{T}}(d)$,  $\underline{P}_{\textrm{T}}(d)$, the achievable $\mathsf{SEP}$, and the modulation efficiency $m$. Upon~inspection of \mbox{Equations \eqref{Receive_SNR} and \eqref{SEP}}, it becomes clear that the $\mathsf{SEP}$ on the forward link can be expressed~as
	\begin{equation}
		\mathsf{SEP} = K_{\textrm{min}}Q\left(\sqrt{d_{\textrm{min}}^2 \frac{P_\textrm{T}(d)}{A_0d^\alpha N_0WN_\textrm{f}M_\textrm{l}}}\right).
	\end{equation}

Although~a high transmit power is desirable for achieving a low SEP and thus reducing the number of retransmissions $N_\textrm{re}$, it may also result in excessively large transmission power for the sensor and increase the EC of each single-direction transmission. Hence, an~optimal transmit power (or in turn the received SNR $\gamma$) exists and is yet to be found for minimizing the EC per successfully transmitted bit. Similarly, the relationship between the EC per successfully transmitted bit and other major parameters mentioned above needs to be studied.

\subsection{Simulation~Results and Discussions}
In the following numerical simulations, we consider the radio links of a WSN designed for smart grid and operating in the industrial--science--medical (ISM)-oriented 2.4 GHz frequency band. Table~\ref{table2} provides a summary of the pertinent simulation parameters, including circuit-related parameter settings as well. Due to the constant envelope characteristic of the CPM signal, a~nonlinear PA with a high $\eta$ value is employed, in~contrast to the general radio architecture. 

\begin{table*}[!htbp]

	\centering
		\caption{A summary of the pertinent simulation parameters.}
		\label{table2}
		\begin{tabular}{c|c}
			\hline
			\textbf{Parameters}&{\textbf{Values}} \\
			\hline
			{Symbol rate for transmitted signals}&{20 ksps} \\
			\hline
			{$L_P$, $L_H$, and $L_L$}&{4, 3, and 30 bytes} \\
			\hline
			{Power spectral density of AWGN at the receiver ($N_0$)}&{$-$174 dBm/Hz} \\
			\hline
			{Noise figure of the RF front-end of the receiver ($N_f$)}&{10 dB} \\
			\hline
			{Equivalent antenna gain ($A_0$)}&{30 dB} \\
			\hline
			{Bandwidth ($W$)}&{20 kHz}\\
			\hline
			{Additional noise ($M_\textrm{l}$)}&{10 dB}\\
			\hline
			{$P_{\textrm{T0}}$}&{15.9 mW} \\
			\hline
			{$P_{R0}$}&{58.2 mW} \\
			\hline
			{$M$}&{2, 4, 8, 16}  \\
			\hline
			{$h$}&{0.75} \\
			\hline
			{$N$}&{ 3} \\
			\hline
			{Path-loss exponent ($\alpha$)}&{3.5}  \\
			\hline
			{Drain efficiency ($\eta$)}&{0.7 for CPM; 0.35 for OQPSK and 16QAM}\\
			\hline
			{Peak-to-average power ratio ($\xi$)}&{0 dB for CPM;  3.5 dB for OQPSK; 6.7 dB for 16QAM}\\
			\hline

		\end{tabular}
	
\end{table*}

\begin{figure}[tbp]
	\centering
	\includegraphics[width=3.5in]{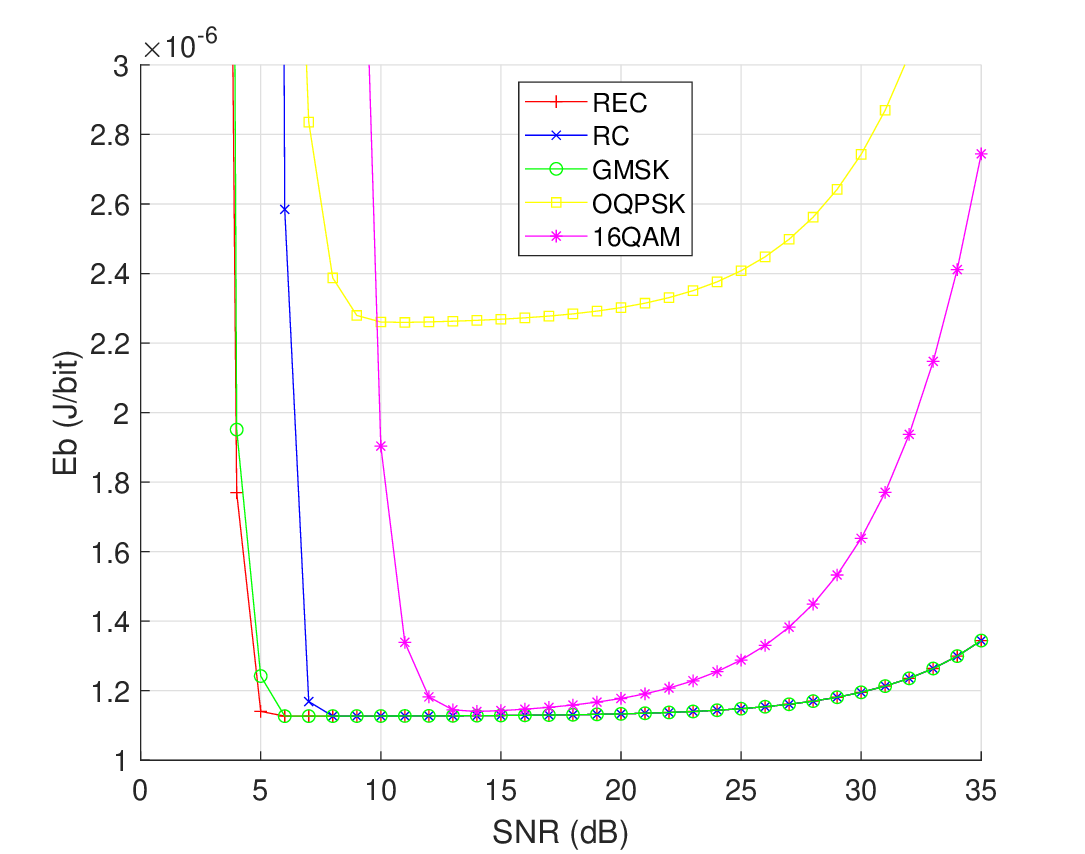}
	\caption{The relationship between the EC per successfully transmitted bit ($E_b$) and the received SNR ($\gamma$) for OQPSK, 16QAM, and the CPM signals with different pulse shaping functions (REC, RC, and GMSK), while assuming $M =16$ and $N=3$ for the three CPM waveforms, as well as $d = 10$ m and the AWGN channel for all the modulation schemes considered.}
	\label{fig3}
\end{figure}

First, let us compare the EC performance of different CPM schemes and of other representative modulation schemes, such as OQPSK and 16QAM, by observing how $E_b$ varies with the received SNR $\gamma$. As~shown in Figure \ref{fig3}, 
for all modulation schemes, the achievable $E_b$ values firstly descend and then increase  with the received SNR. This is because in the low-SNR region, the number of retransmissions plays an important role, while the number of retransmissions is reduced to its minimum under a sufficiently high SNR, and then an even higher SNR means unnecessary energy wastage. In addition, for most SNRs, the achievable~$E_b$ values of REC ($d_{\rm min}^2 = 2.831$, $h= 0.75$ and $N = 3$), RC ($d_{\rm min}^2=8.16596$, $h= 0.75$ and $N = 3$), and GMSK ($d_{\rm min}^2=7.93348$, $h= 0.75$ and $N = 3$) are lower than those of OQPSK and 16QAM, which verifies the advantages of CPM in terms of energy saving. The reason why a higher order CPM scheme has a lower $E_b$ can be explained by referring to Equation~(\ref{eq:EC_per_bit}) as follows: (1) a larger $M$ causes a smaller number of symbols per packet (i.e., the smaller exponent $L/m$) and a smaller $\mathsf{SEP}$ (i.e., the larger base number $1-\mathsf{SEP}$), thus the denominator $L(1-\mathsf{SEP})^{L/m}$ increases with $M$; (2) the variable $T_\textrm{DTA}$ in the numerator becomes smaller when $M$ becomes larger, while all the other terms can be regarded as constants. Furthermore, we can see that the lowest values of $E_b$ achieved by the four modulation schemes having $M = 16$ (i.e., REC, RC, GMSK, and 16QAM) are almost the same, while OQPSK having $M = 4$ achieves the largest $E_b$ in the high-SNR region and the second-largest $E_b$ in the low-SNR region. It is also observed that when the SNR is above a specific  threshold, the three different CPM waveforms exhibit the same EC performance.   

\begin{figure}[tbp]
	\centering
	\includegraphics[width=3.5in]{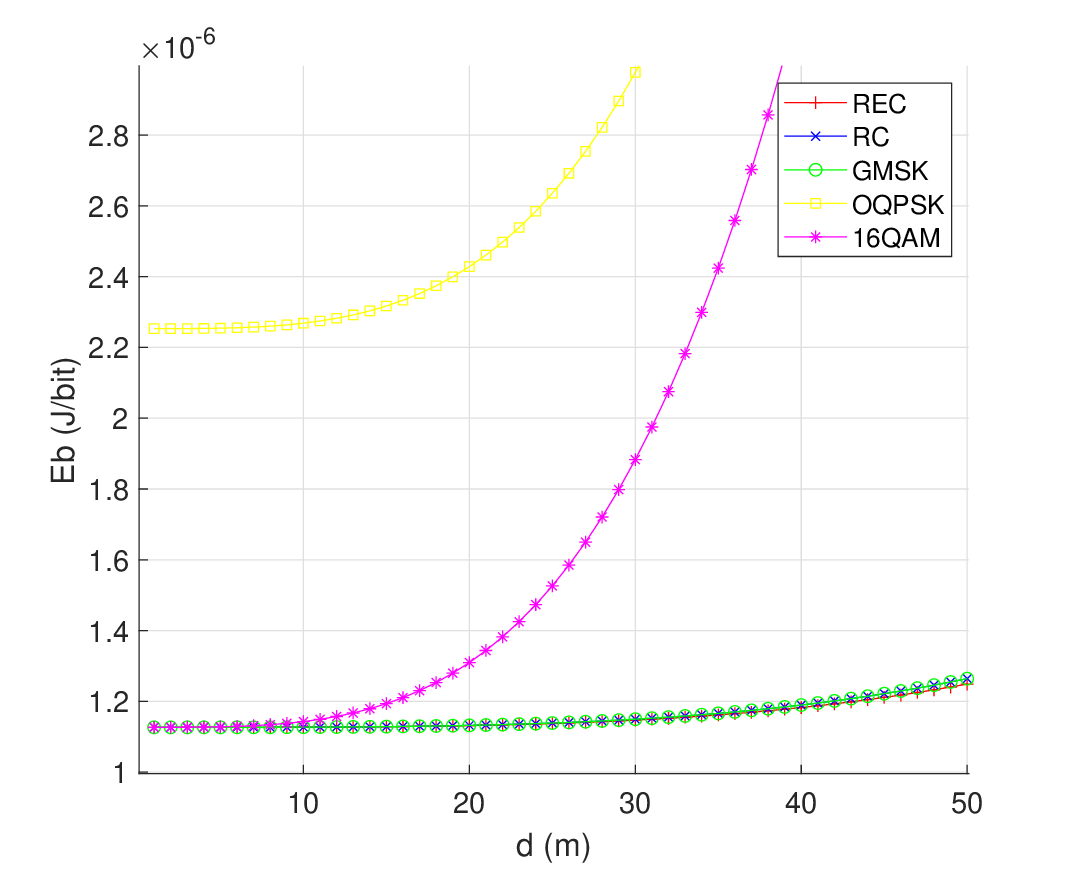}
	\caption{The relationship between the EC per successfully transmitted bit ($E_b$) and the communication distance ($d$) for OQPSK, 16QAM, and the CPM signals with different pulse shaping functions (REC, RC, and GMSK) over the AWGN channel, while assuming $M =16$,  $N=3$ and  $\gamma = 8$ dB for the three CPM waveforms, as well as $\gamma = 15$ dB for OQPSK and 16QAM signals.}
	\label{fig4}
\end{figure}

Figure \ref{fig4} characterizes the relationship between $E_b$ and the communication distance $d$ for OQPSK, 16QAM, and the CPM signals with different pulse shaping functions. We can see that $E_b$ increases with $d$ for all the modulation schemes considered. This is because large distance reduces the received SNR value under a given transmit power, thus degrading the SEP performance and increasing the number of retransmissions. We also see that all the three CPM waveforms considered have the same EC performance curves, which are consistently and significantly better than those of OQPSK and 16QAM when $d$ is sufficiently large. Here, we assume $\gamma = 8$ dB for the three CPM waveforms and $\gamma =15$ dB for OQPSK and 16QAM. The two SNR values are selected according to the results shown in Figure \ref{fig3}, where the three CPM waveforms achieve their optimal $E_b$ at about $\gamma = 8$ dB, while OQPSK and 16QAM achieve their near-optimal $E_b$ at about $\gamma = 15$ dB.

\begin{figure}[tbp]
	\centering
	\includegraphics[width=3.5in]{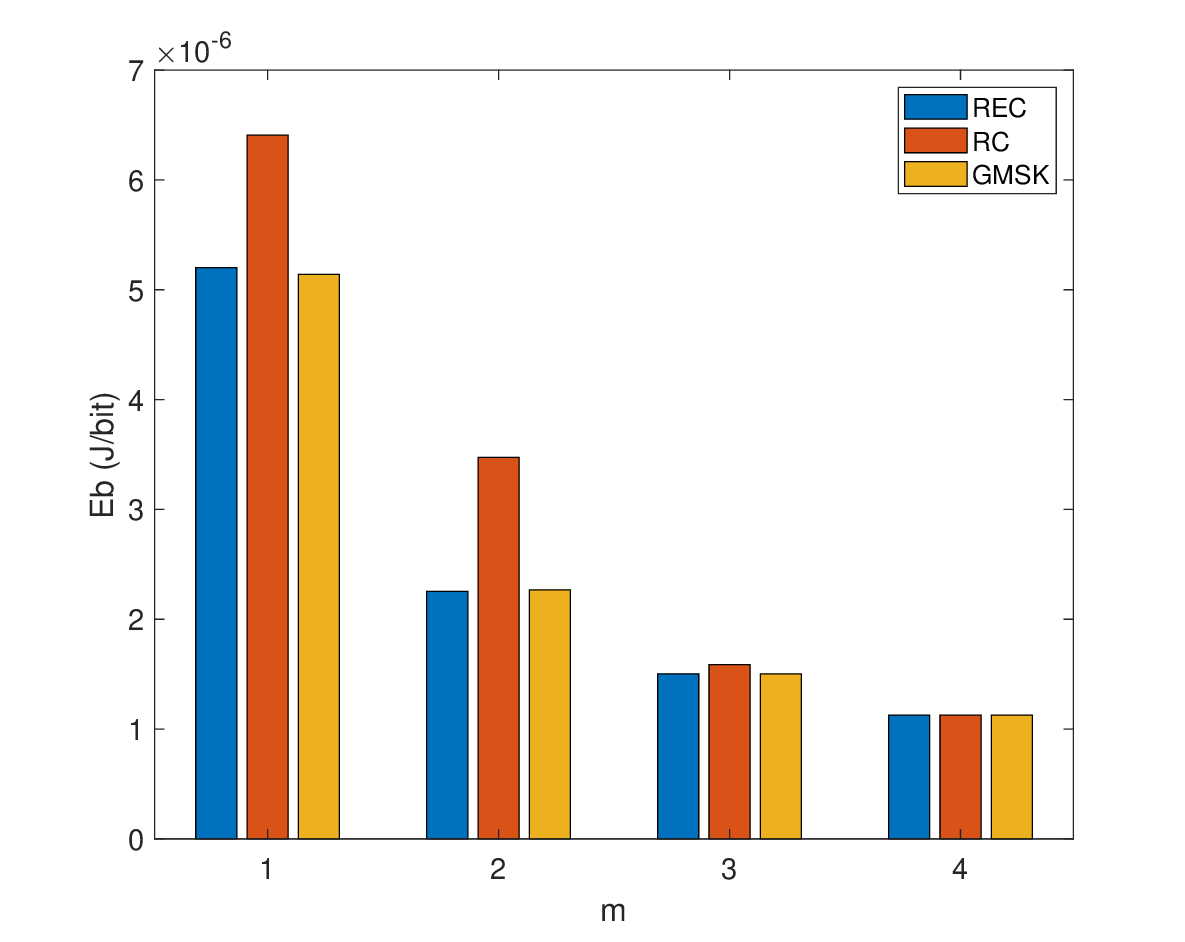}
	\caption{The relationship between the EC per successfully transmitted bit ($E_b$) and the modulation efficiency $m$ for the CPM signals with different pulse shaping functions (REC, RC, and GMSK) over the AWGN channel, while assuming $N=3$, $\gamma = 8$ dB and  $d = 10$ m.}
	\label{fig6}
\end{figure}

Figure \ref{fig6} shows how the $E_b$ values of the three CPM waveforms vary with the modulation efficiency $m$, while assuming $\gamma = 8$ dB and $d=10$ m. It is observed that for each given CPM waveform, the $E_b$ value becomes smaller as $m$ increases. Although the RC-based CPM signaling exhibits the highest $E_b$ when $m = 1, 2, 3$, all the three CPM waveforms achieve almost the same $E_b$ when $m = 4$ (i.e., $M = 16$). This is because the three CPM waveforms achieve almost the same SEP performance when $m = 4$, $\gamma = 8$ dB, and $d=10$ m. These observations are consistent with the results shown in Figure  \ref{fig3}. 

\begin{figure}[tbp]
	\centering
	\includegraphics[width=3.5in]{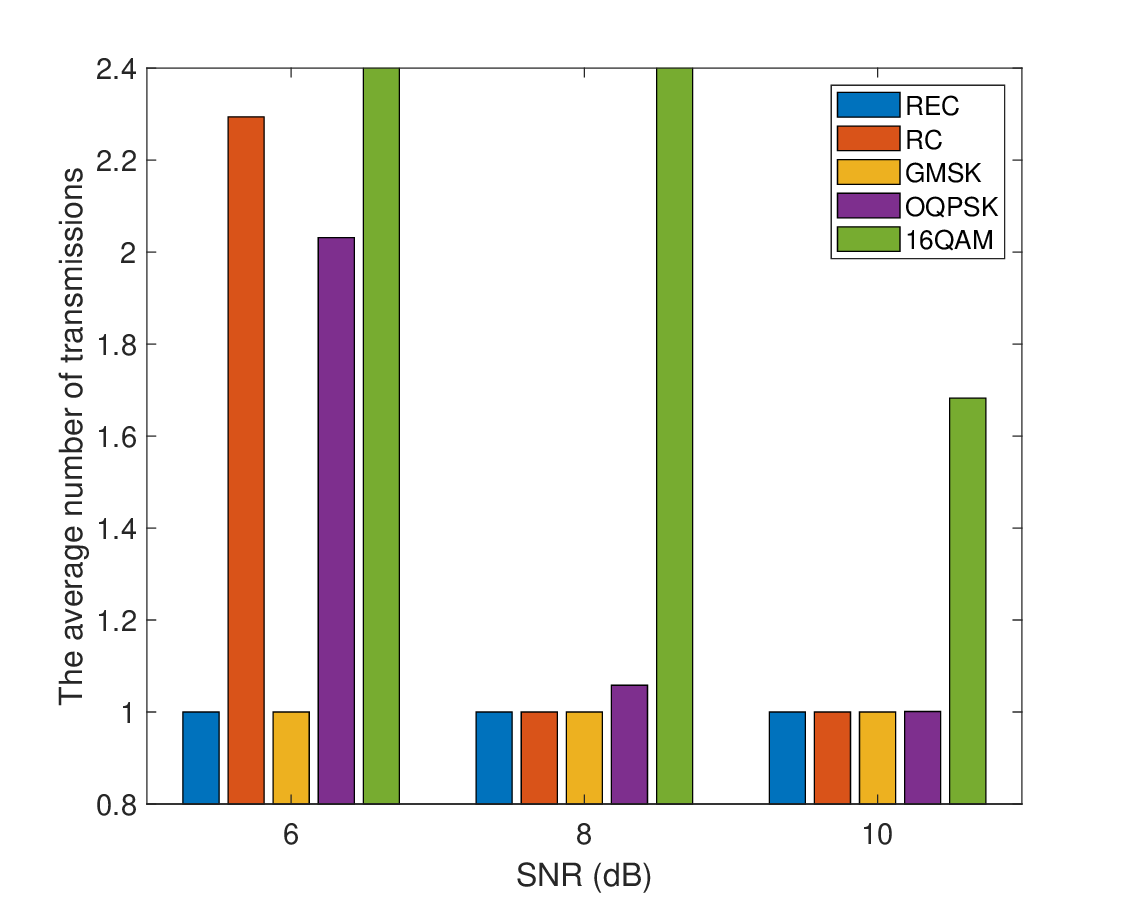}
	\caption{The relationship between the average number of transmissions required for successfully sending a single packet and the received SNR ($\gamma$), while considering OQPSK, 16QAM, and the CPM signals with different pulse shaping functions (REC, RC, and GMSK) over the AWGN channel. Assume that $M =16$ and $N=3$ for the three CPM waveforms, and $d = 10$ m for all the modulation schemes compared.}
	\label{fig7}
\end{figure}

Figure \ref{fig7} demonstrates the relationship between the average number of transmissions $N_\textrm{re}$ required for successfully sending a single packet and the received SNR $\gamma$ over the AWGN channel employing different modulation schemes, including OQPSK, 16QAM, and the three CPM waveforms. Obviously, 16QAM incurs the largest $N_\textrm{re}$, while the REC- and GMSK-based CPM schemes require the smallest $N_\textrm{re}$, under all the three SNR values of \mbox{$6$ dB,} $8$ dB, and $10$ dB. In addition, the OQPSK scheme requires a smaller and a larger $N_\textrm{re}$ than the RC-based CPM scheme under $\gamma = 6$ dB and $\gamma = 8$ dB, respectively.  However, when the received SNR is sufficiently high, e.g., $\gamma = 10$ dB, all the modulation schemes, except 16QAM, require only a single transmission on average for successfully sending \mbox{a packet.}

\section{Conclusions} 

In this paper, the EC characteristics of various CPM schemes are compared with those of OQPSK and 16QAM in the context of WSN-based SG-IoT of beyond 5G. We first propose an EC model for the sensor nodes of WSNs by considering the circuits and a typical communication protocol that relies on ARQ-based retransmissions. Our analytical and simulation results demonstrate that all the CPM schemes based on the pulse shaping functions of REC, RC, and GMSK significantly outperform OQPSK used in the Zigbee standard and 16QAM used in the current 5G standard, in terms of the EC per successfully transmitted bit, $E_b$. We also show that for all the modulation schemes considered, the individual  optimum values of $E_b$ are achieved with the received SNR that is neither too small nor too large. In addition, we show that $E_b$ increases with the communication distance $d$ for all the modulation schemes considered, and decreases with the modulation order $M$ for the three CPM schemes. Overall, it is observed that the REC- and GMSK-based CPM schemes achieve  the best EC performance of all the modulation schemes considered.

\bibliographystyle{IEEEtran}
\bibliography{references}

\end{document}